\documentclass[10pt]{article}
\usepackage[OE]{express}

\begin{document}

\title{Collective atomic-population-inversion and stimulated radiation for
two-component Bose-Einstein condensate in an optical cavity}
\author{Xiuqin Zhao,\authormark{1,2} Ni Liu,\authormark{1,*} and J-Q Liang%
\authormark{1,3}}
\address{\authormark{1}Institute of Theoretical Physics, Shanxi University, Taiyuan, Shanxi 030006,China\\
\authormark{2}Department of Physics, Taiyuan Normal University, Taiyuan, Shanxi 030001,China\\
\authormark{3}jqliang@sxu.edu.cn}

\email{\authormark{*}liuni2011520@sxu.edu.cn}



\begin{abstract}
In this paper we investigate the ground-state properties and related
quantum phase transitions for the two-component Bose-Einstein
condensate in a single-mode optical cavity. Apart from the usual
normal and superradiant phases multi-stable macroscopic quantum
states are realized by means of the spin-coherent-state variational
method. We demonstrate analytically the stimulated radiation from
collective state of atomic population inversion, which does not
exist in the normal Dicke model with single-component atoms. It is
also revealed that the stimulated radiation can be generated only
from one component of atoms and the other remains in the ordinary
superradiant state. However the order of superradiant and
stimulated-radiation states is interchangeable between two
components of atoms by tuning the relative atom-field couplings and
the frequency detuning as well.
\end{abstract}




\section{Introduction}

The Dicke model (DM), which describes an ensemble of two-level atoms
interacting with a single-mode quantized field \cite{Dic54}, plays a
important role in the study of Bose-Einstein condensate (BEC) trapped in an
optical cavity \cite{BGB10,BMB11,RDB13}. It successfully illustrates the
collective and coherent radiations \cite{Dic54}. A second-order phase
transition from a normal phase (NP) to a superradiant phase (SP) was
revealed long ago by increase of the atom-field coupling from weak to strong
regime \cite{WaH73,Hio73,EmB03}.

In order to realize experimentally the quantum phase transition (QPT)
predicted in the DM the collective atom-photon coupling strength ought to be
in the same order of magnitude as the energy level-space of atoms. This
condition is far beyond atom-field coupling region in the conventional
atom-cavity system. Recently the QPT was achieved with a BEC trapped in a
high-finesse optical cavity \cite{BGB10,BMB11,RDB13}. Thus the cavity BEC
has been regarded as a promising platform to explore the exotic many-body
phenomena \cite%
{YCT07,BDR07,RicardoPuebla13,LDM08,SMorrison08,JMZhangW08,CWL08,LaM10,ZPL10,BHS09,SHB10,SND09,SND10,LLM11}%
.

It was recently revealed that an extended DM with multi-mode cavity fields
\cite{Tho77,TSo07} exhibits interesting phenomena, which have important
applications in quantum information and simulation \cite%
{MDM08,NCO09,TOK09,NCi11,YNo11,WMB11,ENZ11,MWB11,EWi13,YSZ13}. Moreover both
abelian and non-abelian gauge potentials \cite{LLe09} are generated in the
two-mode DM, from which the spin-orbit-induced anomalous Hall effect \cite%
{Lar10} is produced as well. With spatial variation of the atom-photon
coupling strength various quantum phases have been predicted such as the
crystallization, spin frustration \cite{GLG10}, spin glass \cite%
{GLG11,SSa11,BSS13} and Nambu-Goldstone mode \cite{FYZ14}. It is shown that
the strong-coupling \cite{WHR13} may lead to the revival of atomic inversion
in a time scale associated with the cavity-field period \cite{KLR}.\ The
optomechanical DM has been also proposed in order to detect the extremely
weak forces \cite{KV08,MG09,FK09,AGHK10,HWAH11,WLL16}.

\ Recently the dynamics induced by atom-pair tunneling \cite%
{LiangJQ,FuLB,ZhangYC} was revealed. The QPT was investigated \cite%
{TimmermansE,PuH} in two-component BECs by means of the semiclassical
approximation\textbf{. }It was demonstrated that\textbf{\ }coupled
two-component BECs in an optical cavity \cite{DongY} display optical \cite%
{DongY}, fluid \cite{SasakiK}, multi-stabilities and capillary instability
\cite{SasakiK,BarankovRA,SchaeybroeckBV,Sasakik}.\textbf{\ }Substantial
many-particle entanglement is also possible in a two-component condensate
with spin degree of freedom \cite{SorensenA,GordonD,MicheliA}\textbf{\ }and
interference between two BECs has been observed \cite{AndrewsMR}.
Particularly, variety of topological excitations is admitted in
multi-component and spinor BECs such as domain walls, abelian and
non-abelian vortices, monopoles, skyrmions, knots, and D-brane solitons \cite%
{EtoM}.

The QPT in DM has been extensively studied \cite%
{Dic54,BGB10,BMB11,CWL08,LLM11,FYZ14,EmaryC,ChenG,Lian1,Lian2} based on
variational method with the help of\textbf{\ }Holstein-Primakoff
transformation \cite{EmB03,CWL08,LLM11,FYZ14,EmaryC,ChenG,NIJL13}\textbf{\ }%
to convert the pseudospin operators into a one-mode bosonic operator in the
thermodynamic limit\textbf{. }The ground-state properties were also revealed
in terms of the catastrophe formalism \cite{GilmoreR}, the dynamic approach
\cite{DimerF,HorakP}, and the spin coherent-state variational method \cite%
{WLL16,Lian1,Lian2,CastannosO,ZLL14,CLS04}, in which both the normal ($%
\Downarrow $) and inverted ($\Uparrow $) pseudospin \cite{NIJL13,KBS10,BMS12}
can be taken into account giving rise to the multi-stable macroscopic
quantum states.

In the present paper, we investigate macroscopic (or collective) quantum
states for two-component BECs in a single-mode optical cavity by means of
the spin coherent variational method in order to reveal the rich structure
of phase diagrams and the related QPTs. Particularly the collective state of
atomic population inversion, namely the inverted pseudospin ($\Uparrow $),
is demonstrated along with the stimulated radiation, which does not exists
in the usual DM.

\section{Collective population inversion and stimulated radiation beyond the
normal and superradiant phases}
\begin{figure}[ht!]
\centering\includegraphics[width=7cm]{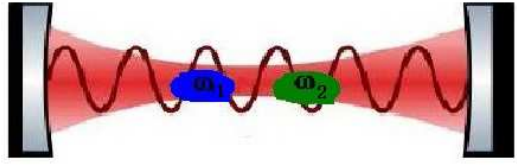}
\caption{Schematic diagram for two ensembles of ultracold atoms (blue and
green) with transition frequencies $\protect\omega _{1}$\textbf{, }$\protect%
\omega _{2}$ in an optical cavity of\textbf{\ }frequency\textbf{\ }$\protect%
\omega $.}
\end{figure}
We consider two ensembles of ultracold atoms, which are coupled
simultaneously to an optical cavity mode of frequency $\omega $ as
depicted in Fig. 1. Effective Hamiltonian of the system has the form
\cite{ABha14} of two-component DM in the unit convention $\hbar =1$,
\begin{eqnarray*}
H &=&\omega a^{\dag }a+\sum_{l=1,2}\omega _{l}J_{lz} \\
&&+\sum_{l=1,2}\frac{g_{l}}{\sqrt{N_{l}}}\left( a^{\dag }+a\right) \left(
J_{l+}+J_{l-}\right).
\end{eqnarray*}%
Where $J_{lz}$ $(J_{l\pm }=J_{lx}\pm iJ_{ly},\quad l=1,2)$ is the collective
pseudospin operator with spin quantum-number $s_{l}$ $=N_{l}/2$. $N_{l}$
denotes the atom number of $l$-th component and $\omega _{l}$ is the atomic
frequency. $a^{\dag }\left( a\right)$ is the photon creation (annihilation)
operator and $g_{l}$ is the atom-field coupling strength.

\section{Spin coherent-state variational method}

In this paper we provide analytic solutions for the macroscopic
quantum state (MQS) for the spin-boson system in terms of the
recently developed spin coherent variational method
\cite{LHC90,RFF99,WLL16,ZLL14}. The meaning of MQS in the present
paper is that the variational wave function is considered as a
product of boson and spin coherent states seen in the followings. We
begin with the partial average of the system Hamiltonian in the
trial wave function $\left\vert \alpha \right\rangle $, which is
assumed as the boson coherent state of cavity mode such that
$a\left\vert \alpha \right\rangle =\alpha \left\vert \alpha
\right\rangle $. After the average in the boson coherent state we
obtain an effective Hamiltonian of the pseudospin operators only,
\begin{equation}
H_{sp}\left( \alpha \right) =\left\langle \alpha \right\vert H\left\vert
\alpha \right\rangle =\omega \alpha ^{\ast }\alpha +\sum_{l=1,2}\omega
_{l}J_{lz} +\sum_{l=1,2}\frac{g_{l}}{\sqrt{N_{l}}}\left( \alpha ^{\ast
}+\alpha \right) \left( J_{l+}+J_{l-}\right),  \label{1}
\end{equation}
which is going to be diagonalized in terms of spin coherent-state
transformation. A spin coherent state can be generated from the maximum
Dicke states $\left\vert s,\pm s\right\rangle $ $(J_{z}$ $\left\vert s,\pm
s\right\rangle =\pm s\left\vert s,\pm s\right\rangle $) with a spin
coherent-state transformation \cite{CLS04,YZLLM96}. For the $l$-th component
pseudospin operator we have two orthogonal coherent states defined by

\[
\left\vert \pm \mathbf{n}_{l}\right\rangle =R(\mathbf{n}_{l})\left\vert
s,\pm s\right\rangle _{l},
\]%
which are called north- and south- pole gauges respectively. As a matter of
fact the spin coherent states are actually the eigenstates of the spin
projection operator $\mathbf{J}_{l}\cdot \mathbf{n}_{l}\left\vert \pm
\mathbf{n}_{l}\right\rangle =\pm j\left\vert \pm \mathbf{n}_{l}\right\rangle
$, where\textbf{\ }$\mathbf{n}_{l}=\left( \sin \theta _{l}\cos \varphi
_{l},\sin \theta _{l}\sin \varphi _{l},\cos \theta _{l}\right) $ is the unit
vector with the directional angles $\theta _{l}$ and $\varphi _{l}$. In the
spin coherent states the spin operators satisfy the minimum uncertainty
relation, for example, $\Delta J_{+}\Delta J_{-}=\left\langle
J_{z}\right\rangle /2$ so that $\left\vert \pm \mathbf{n}\right\rangle $ are
called the MQSs. The unitary operator is explicitly given by
\begin{equation}
R(\mathbf{n}_{l})=e^{\frac{\theta _{l}}{2}(J_{l+}Je^{-i\varphi
_{l}}-J_{l-}e^{i\varphi _{l}})}.  \label{uni}
\end{equation}%
Since pseudospin operators for two components of atoms commute each other,
the entire trail-wave-function is the direct product of two-component spin
coherent states
\[
|\psi _{s}\rangle =\left\vert \pm \mathbf{n}_{1}\right\rangle \left\vert \pm
\mathbf{n}_{2}\right\rangle ,
\]%
which is required to be the energy eigenstate of \ the effective Hamiltonian
of pseudospin operator such that
\begin{equation}
H_{sp}\left( \alpha \right) \left\vert \psi _{s}\right\rangle =E\left(
\alpha \right) \left\vert \psi _{s}\right\rangle .  \label{e}
\end{equation}%
Where%
\begin{equation}
|\psi _{s}\rangle =U\left\vert \pm s\right\rangle _{1}\left\vert \pm
s\right\rangle _{2},  \label{s}
\end{equation}%
with
\[
U=R(\mathbf{n}_{1})R(\mathbf{n}_{2}),
\]%
being the total unitary operator of spin coherent-state transformation. It
is a key point to take into account of both spin coherent states $\left\vert
\pm \mathbf{n}\right\rangle $ for revealing the multi-stable MQSs. Applying
the unitary transformation $U^{\dag }=R^{\dag }(\mathbf{n}_{2})R^{\dag }(%
\mathbf{n}_{1})$ to the energy eigenequation Eq. (\ref{e}) we have
\[
\widetilde{H}_{sp}\left( \alpha \right) \left\vert \pm s\right\rangle
_{1}\left\vert \pm s\right\rangle _{2}=E\left( \alpha \right) \left\vert \pm
s\right\rangle _{1}\left\vert \pm s\right\rangle _{2},
\]%
where
\[
\widetilde{H}_{sp}\left( \alpha \right) =U^{\dag }H_{sp}\left( \alpha
\right) U.
\]%
Under the spin coherent-state transformation the spin operators $J_{lz}$, $%
J_{l+}$, $J_{l-}$ ($l=1$, $2$) become \cite{YZLLM96}

\begin{eqnarray*}
\widetilde{J}_{lz} &=&J_{lz}\cos \theta _{l}+\frac{1}{2}\sin \theta
_{l}\left( J_{l+}e^{-i\varphi _{l}}+J_{l-}e^{i\varphi _{l}}\right),
\end{eqnarray*}%
\begin{eqnarray*}
\widetilde{J}_{l+} &=&J_{l+}\cos ^{2}\frac{\theta _{l}}{2}%
-J_{l-}e^{2i\varphi _{l}}\sin ^{2}\frac{\theta _{l}}{2}-J_{lz}e^{i\varphi
_{l}}\sin \theta _{l},
\end{eqnarray*}%
\begin{eqnarray}
\widetilde{J}_{l-} &=&J_{l-}\cos ^{2}\frac{\theta _{l}}{2}%
-J_{l+}e^{-2i\varphi _{l}}\sin ^{2}\frac{\theta _{l}}{2}-J_{lz}e^{-i\varphi
_{l}}\sin \varphi _{l}.  \label{tr}
\end{eqnarray}%
Then the effective spin Hamiltonian can be diagonalized under the conditions

\begin{eqnarray*}
\frac{\omega _{l}}{2}e^{-i\varphi _{l}}\sin \theta _{l}+\widetilde{g}%
_{l\alpha }\left( \cos ^{2}\frac{\theta _{l}}{2}-e^{-2i\varphi _{l}}\sin ^{2}%
\frac{\theta _{l}}{2}\right) =0,
\end{eqnarray*}
\begin{equation}
\frac{\omega _{1}}{2}e^{i\varphi _{l}}\sin \theta _{l}+\widetilde{g}%
_{l\alpha }\left( \cos ^{2}\frac{\theta _{l}}{2}-e^{2i\varphi _{l}}\sin ^{2}%
\frac{\theta _{l}}{2}\right) =0,  \label{con}
\end{equation}%
from which the angle parameters $\theta _{l}$, $\varphi _{l}$ are determined
in principle. Thus we obtain the energy function

\begin{eqnarray*}
E\left( \alpha \right) =\omega \left\vert \alpha \right\vert ^{2}\pm
\sum_{l=1,2}\frac{N_{l}}{2}A_{l}\left( \alpha ,\theta _{l},\varphi
_{l}\right) ,
\end{eqnarray*}%
where%
\begin{eqnarray*}
A_{l}\left( \alpha ,\theta _{l},\varphi _{l}\right) =\omega _{l}\cos \theta
_{l}-\widetilde{g}_{l\alpha }\left( e^{i\varphi _{l}}+e^{-i\varphi
_{l}}\right) \sin \theta _{l},
\end{eqnarray*}%
with $\widetilde{g}_{l\alpha }=\frac{g_{l}}{\sqrt{N_{l}}}\left( \alpha^{\ast
}+\alpha \right) $. The total trial-wave-function is%
\begin{equation}
\left\vert \psi \right\rangle =\left\vert \alpha \right\rangle \left\vert
\psi _{s}\right\rangle ,  \label{w}
\end{equation}%
and corresponding energies are found as local minima of the energy function $%
E\left( \alpha \right) $, in which the complex eigenvalue of boson coherent
state is parametrized as
\begin{eqnarray*}
\alpha =\gamma e^{i\phi }.
\end{eqnarray*}%
By solving the Eq. (\ref{con}) and eliminating the angle parameters $\theta
_{l},\varphi _{l},\phi $ we derive the scaled-energy as a function of one
variational-parameter $\gamma $ only

\begin{equation}
\frac{E}{\omega }\left( \gamma \right) =\gamma ^{2}\pm \sum_{l=1,2}
\frac{N_{l}}{2}\sqrt{(\frac{\omega _{l}}{\omega })^{2}+\frac{16\gamma ^{2}}{%
N_{l}}(\frac{g_{l}}{\omega })^{2}}.  \label{ener}
\end{equation}%
The local minima of energy function Eq. (\ref{ener}) can be determined in
terms of the variation with respect to the parameter $\gamma $.

\section{Multi-stable states and phase diagram}

In our formalism both the normal ($\Downarrow $) and inverted ($\Uparrow )$
pseudospin states \cite{NIJL13,KBS10} are taken into account to reveal the
multiple stable states. Thus there exist four combinations of two-spin
states labeled by $\downarrow \downarrow $ (both normal spins), $\uparrow
\uparrow $ (both inverted spins), $\downarrow \uparrow $ and $\uparrow
\downarrow $ (first-spin normal, second-spin inverted and vas versa). For
the configuration of both normal spins the dimensionless energy is

\begin{eqnarray*}
\frac{E_{\downarrow \downarrow }\left( \gamma \right) }{\omega }=\gamma
^{2}-\sum_{l=1,2}\frac{N_{l}}{2}\sqrt{(\frac{\omega _{l}}{\omega })^{2}+%
\frac{16\gamma ^{2}}{N_{l}}(\frac{g_{l}}{\omega })^{2}}.
\end{eqnarray*}%
In the following evaluations we assume the equal atom numbers for the two
components that $N_{1}=N_{2}=N/2$. The atomic frequencies are parametrized
according to the cavity frequency $\omega $ and atom-field detuning $\Delta $
\begin{equation}
\omega _{1}=\omega -\Delta, \omega _{2}=\omega +\Delta .  \label{5}
\end{equation}%
The ground-state is obtained from the variation of average energy
\begin{eqnarray*}
\varepsilon _{\downarrow \downarrow }=\frac{E_{\downarrow \downarrow }\left(
\gamma \right) }{N\omega }
\end{eqnarray*}%
with respect to the variational parameter $\gamma $. The energy extremum
condition is found as
\begin{equation}
\frac{\partial \varepsilon _{\downarrow \downarrow }}{\partial \gamma }%
=2\gamma _{\downarrow \downarrow }p_{\downarrow \downarrow }(\gamma
_{\downarrow \downarrow })=0,  \label{ext}
\end{equation}%
where%
\begin{eqnarray*}
p_{\downarrow \downarrow }(\gamma _{\downarrow \downarrow })=1-\sum_{l=1,2}%
\frac{4g_{l}^{2}}{\omega ^{2}F_{l}(\gamma _{\downarrow \downarrow })},
\end{eqnarray*}%
and

\[
F_{l}(\gamma _{\downarrow \downarrow })=\sqrt{(\frac{\omega _{l}}{\omega }%
)^{2}+32(\frac{g_{l}}{\omega })^{2}\frac{\gamma _{\downarrow \downarrow }^{2}%
}{N}}.
\]%
The extremum condition Eq. (\ref{ext}) possesses always a zero photon-number
solution $\gamma _{\downarrow \downarrow }=0$, which is stable if the
second-order derivative of energy function,\bigskip
\[
\frac{\partial ^{2}(\varepsilon _{\downarrow \downarrow }(\gamma
_{\downarrow \downarrow }^{2}=0)}{\partial \gamma ^{2}}=2\left[ 1-\frac{4}{%
\omega }\left( \frac{g_{1}^{2}}{\omega _{1}}+\frac{g_{2}^{2}}{\omega _{2}}%
\right) \right] ,
\]%
is positive. Therefore a phase boundary is determined from $\partial
^{2}(\varepsilon _{\downarrow \downarrow }(\gamma _{\downarrow \downarrow
}^{2}=0)/\partial \gamma ^{2}=0$, which gives rise to the relation of two
critical coupling values
\[
\frac{g_{1,c}^{2}}{\omega _{1}}\mathbf{+}\frac{g_{2,c}^{2}}{\omega _{2}}%
\mathbf{=}\frac{\omega }{4}.
\]%
When\textbf{\ }%
\begin{equation}
\frac{g_{1}^{2}}{\omega _{1}}+\frac{g_{2}^{2}}{\omega _{2}}<\frac{\omega }{4}%
,  \label{a}
\end{equation}%
we have a stable zero photon-number solution, which we call the NP denoted by%
\textbf{\ }$N_{\downarrow \downarrow }$. The energy function for the
configuration $\downarrow \uparrow $ is

\begin{eqnarray*}
\varepsilon _{_{\downarrow \uparrow }}=\frac{\gamma ^{2}}{N}-\frac{1}{4}%
[F_{1}(\gamma _{\downarrow \uparrow })-F_{2}(\gamma _{\downarrow \uparrow })]
\end{eqnarray*}%
The energy extremum condition $\partial \varepsilon _{_{\downarrow \uparrow
}}/\partial \gamma =2\gamma _{\downarrow \uparrow }p_{\downarrow \uparrow
}(\gamma _{\downarrow \uparrow })=0$ with

\begin{eqnarray*}
p_{\downarrow \uparrow }(\gamma _{\downarrow \uparrow })=1-\frac{4}{\omega
^{2}}\left[ \frac{g_{1}^{2}}{F_{1}(\gamma _{\downarrow \uparrow })}-\frac{%
g_{2}^{2}}{F_{2}(\gamma _{\downarrow \uparrow })}\right] ,
\end{eqnarray*}%
has the zero photon-number solution, which is stable when the second-order
derivative

\begin{eqnarray*}
\frac{\partial ^{2}\left( \varepsilon _{_{\downarrow \uparrow }}(\gamma
_{\downarrow \uparrow }^{2}=0)\right) }{\partial \gamma ^{2}}=\frac{2}{N}%
\left[ 1-\frac{4}{\omega }\left( \frac{g_{1}^{2}}{\omega _{1}}-\frac{%
g_{2}^{2}}{\omega _{2}}\right) \right]
\end{eqnarray*}%
is positive. Thus we have the NP (denoted by $N_{\downarrow \uparrow }$ )
region when
\begin{equation}
\frac{g_{1}^{2}}{\omega _{1}}-\frac{g_{2}^{2}}{\omega _{2}}<\frac{\omega }{4}%
.  \label{b}
\end{equation}%
\ Correspondingly for the configuration $\uparrow \downarrow $ the energy
function is

\begin{eqnarray*}
\varepsilon _{_{\uparrow \downarrow }}=\frac{\gamma ^{2}}{N}+\frac{1}{4}%
\left[ F_{1}(\gamma )-F_{2}(\gamma )\right] .
\end{eqnarray*}%
The energy extremum condition is $\partial \varepsilon _{_{\uparrow
\downarrow }}/\partial \gamma =\gamma _{\uparrow \downarrow }p_{\uparrow
\downarrow }(\gamma _{\uparrow \downarrow })=0$ with

\begin{eqnarray*}
p_{\uparrow \downarrow }(\gamma _{\uparrow \downarrow })=1+\frac{4}{\omega
^{2}}\left( \frac{g_{1}^{2}}{F_{1}(\gamma _{\uparrow \downarrow })}-\frac{%
g_{2}^{2}}{F_{2}(\gamma _{\uparrow \downarrow })}\right) .
\end{eqnarray*}
Again the stable zero photon-number solution denoted by $N_{\uparrow
\downarrow }$ requires

\begin{equation}
\frac{g_{2}^{2}}{\omega _{2}}-\frac{g_{1}^{2}}{\omega _{1}}<\frac{\omega }{4}%
.  \label{c}
\end{equation}%
The energy function for the configuration $\uparrow \uparrow $ is

\begin{eqnarray*}
\varepsilon _{\uparrow \uparrow }=\frac{E_{\uparrow \uparrow }\left( \gamma
\right) }{\omega N}=\frac{\gamma ^{2}}{N}+\frac{1}{4}\sum_{l=1,2}F_{l}(%
\gamma ).
\end{eqnarray*}%
The extremum condition is

\begin{eqnarray*}
\frac{\partial (\varepsilon _{\uparrow \uparrow })}{\partial \gamma }%
=2\gamma _{\uparrow \uparrow }p_{\uparrow \uparrow }(\gamma _{\uparrow
\uparrow })=0,
\end{eqnarray*}%
with

\begin{eqnarray*}
p_{\uparrow \uparrow }(\gamma _{\uparrow \uparrow })=1+\sum_{l=1,2}\frac{%
4g_{l}^{2}}{\omega ^{2}F_{l}(\gamma _{\uparrow \uparrow })}.
\end{eqnarray*}%
The zero photon-number solution is stable denoted by $N_{\uparrow \uparrow }$
since the second-order derivative

\[
\frac{\partial ^{2}\varepsilon _{\uparrow \uparrow }\left( \gamma _{\uparrow
\uparrow }^{2}=0\right) }{\partial \gamma ^{2}}=\frac{2}{N}\left[ 1+\frac{4}{%
\omega }\left( \frac{g_{1}^{2}}{\omega _{1}}+\frac{g_{2}^{2}}{\omega _{2}}%
\right) \right] >0,
\]%
is always positive. The nonzero-photon solution can be obtained from the
extremum condition.%
\begin{equation}
p_{k}(\gamma _{sk})=0  \label{sp}
\end{equation}%
for the four configurations $k=\downarrow \downarrow ,\downarrow \uparrow
,\uparrow \downarrow ,\uparrow \uparrow $. The extremum condition Eq. (\ref%
{sp}) is able to be solved numerically. We display in Fig. 2(a) the stable
nonzero photon solutions $\gamma _{sk}$, which are called the superradiant
states, and the corresponding energies $\varepsilon (\gamma _{sk})$ as shown
in Fig. 2(b) for $k=\downarrow \downarrow $ (black line)$,\downarrow
\uparrow $ (olive line)$,\uparrow \downarrow $ (blue line) respectively. For
the dimensionless coupling $g_{2}/\omega =0.2$ [Figs. 2(a1) and 2(b1)] both
solutions $\gamma _{s\downarrow \downarrow }$ and $\gamma _{s\downarrow
\uparrow }$ of the extremum equation are stable with a positive sloop [Fig.
2(a1)], namely a positive second-order derivative of the energy function
with respect the variation parameter $\gamma $. The corresponding energies
are local minima [Fig. 2(b1)]. $\gamma _{s\downarrow \uparrow }$ indicates
the solution of stimulated radiation from the state of atomic population
inversion for the second-component of atoms. Increasing the coupling
strength to $g_{2}/\omega =0.4$ [Figs. 2(a2) and 2(b2)] and $0.7$ we have
only one stable solution $\gamma _{s\downarrow \downarrow }$. While the two
stable solutions appear again for $g_{2}/\omega =0.9$ [Figs. 2(a4) and
2(b4)]. It is interesting to see a fact that the the stimulated radiation
becomes the first-component of atoms i.e. $\gamma _{s\uparrow \downarrow }$.
The superradiant states are denoted respectively by $S_{\downarrow
\downarrow }$,\textbf{\ }$S_{\uparrow \downarrow }$\textbf{\ }and\textbf{\ }$%
S_{\downarrow \uparrow }$ in the following phase diagrams.
\begin{figure}[th]
\centering\includegraphics[width=5cm]{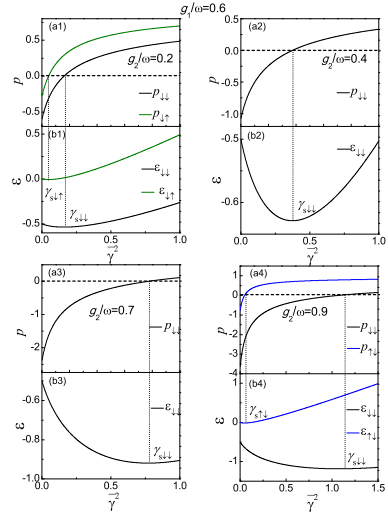}
\caption{Graphical solutions of the extremum equation $p_{k}(\protect\gamma %
_{sk})=0$\ for $k=\downarrow \downarrow $ (black line), $k_{\downarrow
\uparrow }$\ (olive line), and $k=\uparrow \downarrow $\ (blue line) with $%
g_{1}/\protect\omega =0.6$\textbf{\ }and $g_{2}$\textbf{/}$\protect\omega %
=0.2$ (a1), $0.4$ (a2), $0.7$ (a3), $0.9$ (a4).\textbf{\ }The\textbf{\ }%
corresponding average energy curves $\protect\varepsilon $ are plotted in
the lower panel (b1-b4). $\bar{\protect\gamma}^{2}$\ $=\protect\gamma ^{2}/N$
denotes the mean photon number.}
\end{figure}
\begin{figure}[ht!]
\centering\includegraphics[width=5cm]{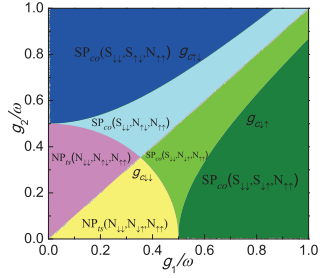}
\caption{Phase diagram in the resonance\textbf{\ }condition\textbf{\ }$%
\protect\omega _{1}=\protect\omega _{2}=\protect\omega $. The notations $%
NP_{ts}\left( N_{\downarrow \downarrow },N_{\uparrow \downarrow }\mathbf{,}%
N_{\downarrow \downarrow }\right)$ and $NP_{ts}\left( N_{\downarrow
\downarrow },N_{\downarrow \uparrow }\mathbf{,}N_{\downarrow \downarrow
}\right) $ mean the NP with triple states, in which\textbf{\ }$N_{\downarrow
\downarrow }$ is the ground state. $SP_{co}\left( S_{\downarrow \downarrow
},N_{\uparrow \downarrow }\mathbf{,}N_{\uparrow \uparrow }\right)$ [ $%
SP_{co}\left( S_{\downarrow \downarrow },N_{\downarrow \uparrow }\mathbf{,}%
N_{\uparrow \uparrow }\right) $]\ means the SP characterized by the ground
state $S_{\downarrow \downarrow }$, which coexists with $N_{\uparrow
\downarrow }$ ( $N_{\downarrow \uparrow }$) and $N_{\uparrow \uparrow }$.%
\textbf{\ }$SP_{co}\left( S_{\downarrow \downarrow },S_{\uparrow \downarrow }%
\mathbf{,}N_{\uparrow \uparrow }\right)$ [ $SP_{co}\left( S_{\downarrow
\downarrow },S_{\downarrow \uparrow }\mathbf{,}N_{\uparrow \uparrow }\right)
$] is also the coexisting SP, in which the first excited-state is a
superradiant state \textbf{\ }$S_{\uparrow \downarrow }$ \textbf{( }$%
S_{\downarrow \uparrow }$\textbf{) }. }
\end{figure}
\begin{figure}[ht!]
\centering\includegraphics[width=5
cm]{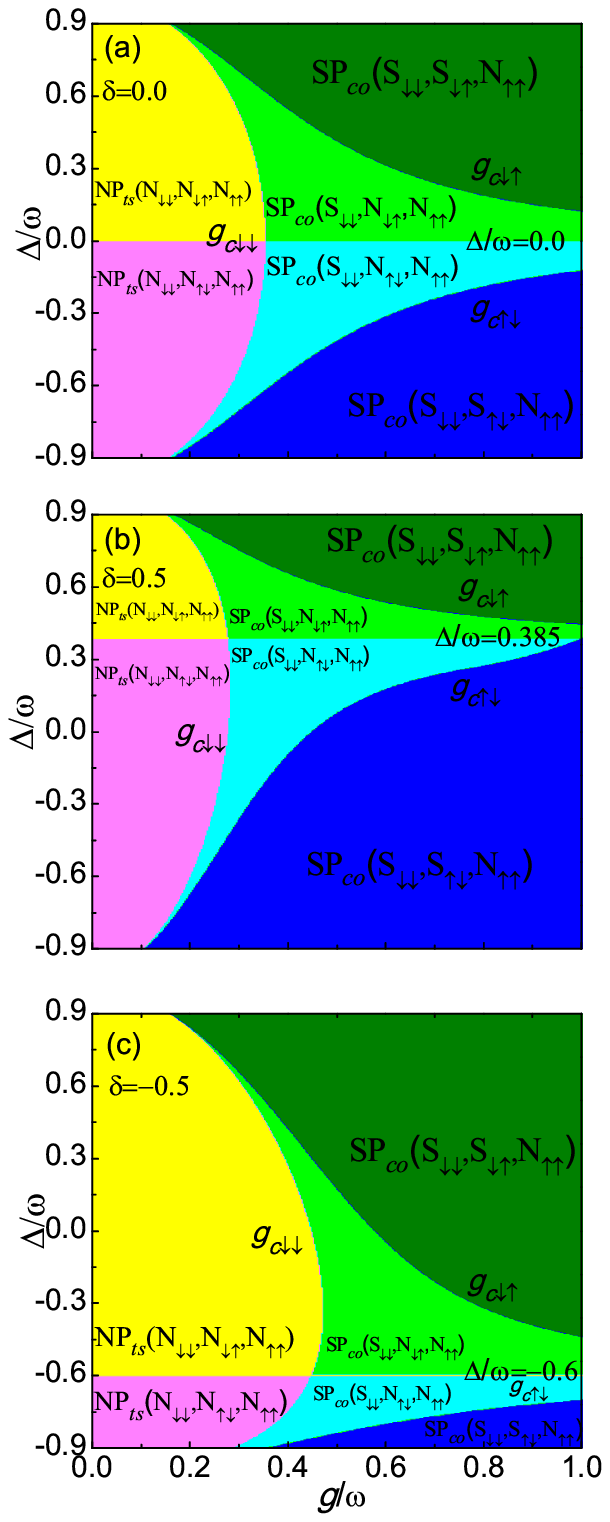} \caption{Phase
diagram in $g$-$\Delta $\ space with the atom-photon coupling
parameter $\protect\delta =0$ (a), $\protect\delta =0.5$ (b), and $\protect%
\delta =-0.5$ (c).\textbf{\ }\ The boundary line, which separates the
regions with different first-excited-states ($N_{\downarrow \uparrow }$, $%
S_{\downarrow \uparrow }$ and $N_{\uparrow \downarrow }$, $S_{\uparrow
\downarrow }$), moves upward and downward respectively for $\protect\delta %
=0.5$ (b) , $-0.5$ (c). }
\end{figure}
The new observation with the spin coherent-state variational-method is that
besides the ground states we also obtained the stable MQSs of higher
energies. Fig. 3 depicts the phase diagram\textbf{\ }in $g_{1}$-$g_{2}$\
plane\textbf{\ }with the resonance condition\textbf{\ }$\omega _{1}=\omega
_{2}=\omega $.\textbf{\ }The phase boundaries $g_{c\downarrow \downarrow }$ $%
g_{c\downarrow \uparrow }$ $g_{c\uparrow \downarrow }$ are determined from
the following three relations respectively%
\begin{eqnarray*}
\mathbf{\ }g_{2}=\frac{1}{2}\sqrt{1-\left( \frac{2g_{1}}{\omega }\right) ^{2}%
},
\end{eqnarray*}%
\begin{eqnarray*}
g_{2}=\frac{1}{2}\sqrt{\left( \frac{2g_{1}}{\omega }\right) ^{2}+1},
\end{eqnarray*}%
\begin{eqnarray*}
g_{2}=\frac{1}{2}\sqrt{\left( \frac{2g_{1}}{\omega }\right) ^{2}-1}.
\end{eqnarray*}%
In the region denoted by $NP_{ts}$ (bounded by the critical line $%
g_{c\downarrow \downarrow }$) there exist triple zero-photon states, in which%
\textbf{\ }$N_{\downarrow \downarrow }$ with lowest energy is the ground
state. This region is separated into two areas (pink and yellow) with only
one state difference that the state $N_{\downarrow \uparrow }$ in one area
is replaced by $N_{\uparrow \downarrow }$ in the other. We see the
simultaneous spin-flip from the state $N_{\downarrow \uparrow }$ to $%
N_{\uparrow \downarrow }$ by adjusting the ratio of two coupling constants
from $g_{2}/g_{1}<1$ (yellow region) to $g_{2}/g_{1}>1$ (pink region). The
notation, for example, $SP_{co}(S_{\downarrow \downarrow },N_{\uparrow
\downarrow },N_{\uparrow \uparrow })$ (cyan area) means the SP region
characterized by the superradiant ground-state $S_{\downarrow \downarrow }$
coexisting with the first ($N_{\uparrow \downarrow }$) and second ($%
N_{\uparrow \uparrow }$) excited states of zero photons). The phase diagram
is symmetric with respect to the line $g_{2}/g_{1}=1$, which separates the
SP region to two areas. Below the symmetric line (green area) only the first
excited state is changed to $N_{\downarrow \uparrow }$ by the
coupling-variation induced spin flip. The critical line $g_{c\uparrow
\downarrow }$ is a boundary, above which the first excited state becomes
surperradiant state $S_{\uparrow \downarrow }$ (cyan region) in the upper
area of the symmetric line. While $g_{c\downarrow \uparrow }$ is the
corresponding boundary for the first excited states $N_{\downarrow \uparrow
} $ and $S_{\downarrow \uparrow }$ (olive area). The superradiant states $%
S_{\uparrow \downarrow }$, $S_{\downarrow \uparrow }$, which are new
observation for the two-component BECs, are seen to be the stimulated
radiation from the higher-energy atomic levels. The stable population
inversion state $N_{\uparrow \uparrow }$ for both components exists in the
whole region.\textbf{\ }The multi-stable MQSs observed in this paper agree
with the dynamic study of nonequilibrium QPTs \cite{KBS10,NIJL13}.

\textbf{\ }We now consider the phase diagram for the atom-field detuning $%
\omega _{1}=\omega -\Delta $ and $\omega _{2}=\omega +\Delta $ with $\Delta
\in \left[ -0.9,0.9\right] $ and the atom-field coupling imbalance parameter
$\delta $ given by \textbf{\ }%
\begin{equation}
g_{1}=g,g_{2}=(1+\delta )g.  \label{6}
\end{equation}%
Substituting atom-field coupling Eq. (\ref{6}) into the corresponding
ground-state energy function we obtain the phase diagram of $g$-$\Delta $
space displayed in Fig. 4 for the imbalance parameter $\delta =0$ [Fig.
4(a)], $0.5$ [Fig. 4(b)], $-0.5$ [Fig. 4(c)]. The phase boundary line $%
g_{c\downarrow \downarrow }$ for the normal state $N_{\downarrow \downarrow
} $ is found from Eq. (\ref{a})

\begin{equation}
g_{c_{\downarrow \downarrow }}=\frac{1}{2}\sqrt{\frac{(\omega ^{2}-\Delta
^{2})}{\omega \left[ 2\omega +(\omega -\Delta )\left( 2\delta +\delta
^{2}\right) \right] }},  \label{a1}
\end{equation}%
The phase diagram for $\delta =0$ as depicted in Fig. 4(a) is symmetric with
respect to the horizontal line $\Delta =0$. The triple-state NP region
denoted by\textbf{\ }$NP_{ts}(N_{\downarrow \downarrow },N_{\downarrow
\uparrow },N_{\uparrow \uparrow }\mathbf{)}$ (yellow) and $%
NP_{ts}(N_{\downarrow \downarrow },N_{\uparrow \downarrow },N_{\uparrow
\uparrow }\mathbf{)}$ (pink) is located on the left-hand side of the
critical line $g_{c\downarrow \downarrow }$, which\ shifts towards the lower
value direction of the atom-field coupling $g$ \cite{LLM11,NIJL13} with the
increase of absolute value of detuning $\left\vert \Delta \right\vert $ seen
from Fig. 4(a). $SP_{co}\left( S_{\downarrow \downarrow },N_{\downarrow
\uparrow }\mathbf{,}N_{\uparrow \uparrow }\right) $ (green region) and $%
SP_{co}\left( S_{\downarrow \downarrow },N_{\uparrow \downarrow }\mathbf{,}%
N_{\uparrow \uparrow }\right) $ (cyan) denote the SP characterized by the
ground-state $S_{\downarrow \downarrow }$\ coexisting with the normal states
$N_{\downarrow \uparrow }$, $N_{\uparrow \downarrow }$ and $N_{\uparrow
\uparrow }$ respectively. The QPT from the NP of ground-state $N_{\downarrow
\downarrow }$\ to the SP of ground-state $S_{\downarrow \downarrow }$\ by
the variation of atom-field coupling $g$ is the standard DM type for the
fixed atom-field detuning $\Delta $. The phase boundary lines $%
g_{c\downarrow \uparrow }$, $g_{c\uparrow \downarrow }$, which separate the
states $S_{\downarrow \uparrow }$ and $S_{\uparrow \downarrow }$, are
respectively\ determined from Eqs. (\ref{b}, \ref{c})\textbf{\ }

\begin{equation}
g_{c\downarrow \uparrow }=\frac{1}{2}\sqrt{\frac{(\omega ^{2}-\Delta ^{2})}{%
\omega \left[ (\omega +\Delta )-(\omega -\Delta )(1+\delta )^{2}\right] }}=%
\frac{1}{2}\sqrt{\frac{(\omega ^{2}-\Delta ^{2})}{\omega \left[ 2\Delta
-(\omega -\Delta )\left( 2\delta +\delta ^{2}\right) \right] }},  \label{b1}
\end{equation}
\textbf{\ }and%
\begin{equation}
g_{c\uparrow \downarrow }=\frac{1}{2}\sqrt{\frac{(\omega ^{2}-\Delta ^{2})}{%
\omega \left[ (\omega -\Delta )(1+\delta )^{2}-(\omega +\Delta )\right] }}=%
\frac{1}{2}\sqrt{\frac{(\omega ^{2}-\Delta ^{2})}{\omega \left[ (\omega
-\Delta )\left( 2\delta +\delta ^{2}\right) -2\Delta \right] }}.  \label{c1}
\end{equation}
The superradiant region denoted by $SP_{co}\left( S_{\downarrow \downarrow
},S_{\downarrow \uparrow }\mathbf{,}N_{\uparrow \uparrow }\right) $ (olive
area) is above the the critical line $g_{c\downarrow \uparrow }$, while $%
SP_{co}\left( S_{\downarrow \downarrow },S_{\uparrow \downarrow }\mathbf{,}%
N_{\uparrow \uparrow }\right) $ (blue) is located below the critical line $%
g_{c\uparrow \downarrow }$. We see that the second excited-state varies from
the normal state $N_{\downarrow \uparrow }$ to the superradiant state $%
S_{\downarrow \uparrow }$ by the increase of detuning $\Delta $. The
difference of upper and lower half-plane of the phase diagram is made only
by the first excited-states $N_{\downarrow \uparrow }$, $S_{\downarrow
\uparrow }$ and $N_{\uparrow \downarrow }$, $S_{\uparrow \downarrow }$ with
the interchange of spin polarizations between two components. This boundary
line, which separates the regions with different first excited-states, moves
upward and downward respectively for $\delta =0.5$ [Fig. 4(b)], $-0.5$ [Fig.
4(c)].

\section{Mean photon number, atomic population and average energy from
viewpoint of phase transition}

The mean photon numbers in the states $N_{\downarrow \downarrow }$
and $S_{\downarrow \downarrow }$\ can be evaluated directly from the
average
of photon number-operator in the corresponding wave functions\textbf{\ }$%
|\psi \rangle =\left\vert \alpha \right\rangle \left\vert \psi
_{s}\right\rangle $ in Eq. (\ref{w}) with spin-state $|\psi
_{s}(-s,-s)\rangle =U\left\vert -s\right\rangle _{1}\left\vert
-s\right\rangle _{2}$\textbf{\ }given in Eq. (\ref{s}). The result
is obviously
\begin{eqnarray*}
n_{p}(\downarrow \downarrow )=\frac{\left\langle \alpha |a^{\dag }a|\alpha
\right\rangle }{N}=\left\{
\begin{array}{r}
0,\qquad\ g<g_{c\downarrow \downarrow }, \\
\frac{\gamma _{\downarrow \downarrow }^{2}}{N}, \qquad g>g_{c\downarrow
\downarrow }.%
\end{array}%
\right..
\end{eqnarray*}%
While the atomic population imbalance becomes%
\begin{eqnarray*}
\Delta n_{a}\left( \downarrow \downarrow \right) =\frac{\left\langle \psi
_{s}(-s,-s)|(J_{1z}+J_{2z})|\psi _{s}(-s,-s)\right\rangle }{N} =-\frac{1}{4}%
\sum_{l=1,2}\frac{\omega _{l}}{\omega F_{l}(\gamma _{\downarrow \downarrow })%
},
\end{eqnarray*}%
which reduces to the well-known standard Dicke-model value
\begin{eqnarray*}
\Delta n_{a}(\downarrow \downarrow )=-\frac{1}{2},
\end{eqnarray*}%
at the critical line $g_{c\downarrow \downarrow }$ and also the NP state $%
N_{\downarrow \downarrow }$.
The average energy in ground states $N_{\downarrow \downarrow }$ and $%
S_{\downarrow \downarrow }$ is given by
\begin{eqnarray*}
\varepsilon _{_{\downarrow \downarrow }}=\left\{
\begin{array}{r}
-0.5, g<g_{c\downarrow \downarrow }, \\
\frac{\gamma _{\downarrow \downarrow }^{2}}{N}-\frac{1}{4}%
\sum\limits_{l=1,2}F_{l}(\gamma _{\downarrow \downarrow }), g>g_{c\downarrow
\downarrow }.%
\end{array}%
\right. .
\end{eqnarray*}%
\begin{figure}[tp]
\centering\includegraphics[width=5cm]{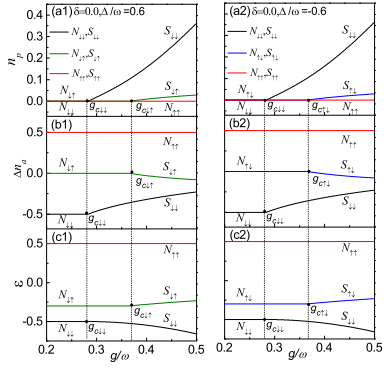}\caption{Variations
of the average photon number $n_{p}$ (a), atom
population imbalance $\Delta n_{a}$ (b), and average energy $\protect%
\varepsilon $\ (c) with respect to the coupling constant $g=$
$g_{1}=g_{2}$ in the atom-field frequency detuning $\Delta =0.6$ (1)
and $\Delta =-0.6$ (2).}
\end{figure}
For the states $N_{k}$ and $S_{k}$ with opposite spin-polarizations $%
k=\downarrow \uparrow $,$\uparrow \downarrow $ the average photon number is
\
\begin{eqnarray*}
n_{p}(N_{k})=0;\qquad n_{p}(S_{k})=\frac{\gamma _{k}^{2}}{N}.
\end{eqnarray*}%
The atomic population imbalance becomes%
\begin{eqnarray*}
\Delta n_{a}(N_{k})=0
\end{eqnarray*}%
for the zero-photon states $N_{k}$. While the atomic population imbalance
for the superradiant states $S_{k}$ is seen to be
\begin{eqnarray*}
\Delta n_{a}(S_{\downarrow \uparrow })=\frac{1}{4\omega }\left[ -\frac{%
\omega _{1}}{F_{1}(\gamma _{\downarrow \uparrow })}+\frac{\omega _{2}}{%
F_{2}(\gamma _{\downarrow \uparrow })}\right] ,
\end{eqnarray*}

\[
\Delta n_{a}(S_{\uparrow \downarrow })=\frac{1}{4\omega }\left[ \frac{\omega
_{1}}{F_{1}(\gamma _{\uparrow \downarrow })}-\frac{\omega _{2}}{F_{2}(\gamma
_{\uparrow \downarrow })}\right] .
\]%
The average energies $\varepsilon _{k}(S_{k})$ of the superradiant states $%
S_{k}$ for $k=\downarrow \uparrow $,$\uparrow \downarrow $ can be obtained
from the energy functions with the corresponding solutions $\gamma _{k}$,
which lead to $\varepsilon _{k}(N_{k})=0$. For the inverted-spin state of
zero photon the atomic population imbalance is $\Delta n_{a}(N_{\uparrow
\uparrow })=0.5$ and the average energy is found as
\[
\varepsilon (N_{\uparrow \uparrow })=\frac{1}{4\omega }(\omega _{1}+\omega
_{2}).
\]%
The stable nonzero-photon state does not exists for this
configuration of both inverted spins. The average photon number
$n_{p}$, atomic population imbalance $\Delta n_{a}$, and the average
energy $\varepsilon $ are plotted in Fig. 5 as functions of the
atom-field coupling strength $g$ in the red and blue detuning
$\Delta =\pm 0.6$ with $\delta =0$. Below the critical point
$g_{c\downarrow \downarrow }$ we have triple stable (zero-photon)
states denoted by $NP_{ts}\left( N_{\downarrow \downarrow
},N_{\downarrow \uparrow }\mathbf{,}N_{\uparrow \uparrow }\right) $
[or $NP_{ts}\left( N_{\downarrow \downarrow },N_{\uparrow \downarrow
}\mathbf{,}N_{\uparrow \uparrow }\right) $], in which $N_{\downarrow
\downarrow }$ (black line) is
the ground state with lowest energy. Between the critical points $%
g_{c\downarrow \downarrow }$\ and $g_{c\downarrow \uparrow }$ (or $%
g_{c\uparrow \downarrow }$) the superradiant ground-state $S_{\downarrow
\downarrow }$ (black line) coexists with the states\ $N_{\downarrow \uparrow
}$ [olive lines in Figs. 5(a1)-5(c1)], or $N_{\uparrow \downarrow }$ [blue
lines in Figs. 5(a2)-5(c2)], and $N_{\uparrow \uparrow }$ (red lines). The
QPT from the NP ($N_{\downarrow \downarrow }$)\ to the SP ($S_{\downarrow
\downarrow }$)\ is the standard DM type, which takes place at the critical
point $g_{c\downarrow \downarrow }$. From Fig. 5(c1) we see that the states
\ $N_{\downarrow \uparrow }$ and\ $S_{\downarrow \uparrow }$ (olive lines)
of opposite spin-polarizations are the first excited-states in the case $%
\Delta =6$. While the states $N_{\uparrow \downarrow }$\ and\ $S_{\uparrow
\downarrow }$ with interchange of the spin polarizations between two
components become the first excited states seen from Fig. 5(c2) (blue lines)
for the negative detuning $\Delta =-6$. We observe for the first time the
phase transition at the critical point $g_{c\downarrow \uparrow }$ $%
(g_{c\uparrow \downarrow }$) from the normal state $N_{\downarrow \uparrow }$
( $N_{\uparrow \downarrow }$) to the suprradiant state $S_{\downarrow
\uparrow }$ ($S_{\uparrow \downarrow }$), which is the stimulated radiation
from the collective states of atomic population inversion for one component
of BECs seen from Figs. 5 and 6.\ The ground state does not change at the
critical point $g_{c\downarrow \uparrow }$ (or $g_{c\uparrow \downarrow }$),
which separates the normal state $N_{\downarrow \uparrow }$ (or $N_{\uparrow
\downarrow }$) and the superradiant one $S_{\downarrow \uparrow }$ (or $%
S_{\uparrow \downarrow }$), which are the collective excited-states of the
system. For the given frequency detuning $\Delta =\pm 0.6$ (Fig. 5) the
critical points can be evaluated precisely for Eq. (\ref{a1}, \ref{b1}, \ref%
{c1}), $g_{c\downarrow \downarrow }=\sqrt{2}/5=0.2828$ and $g_{c\downarrow
\uparrow }=g_{c\uparrow \downarrow }\mathbf{=}\sqrt{2/15}=0.365148$. The
normal state $N_{\uparrow \uparrow }$ (red line) of atomic population
inversion for both components does not involve in radiation process.
\begin{figure}[th]
\centering\includegraphics[width=5cm]{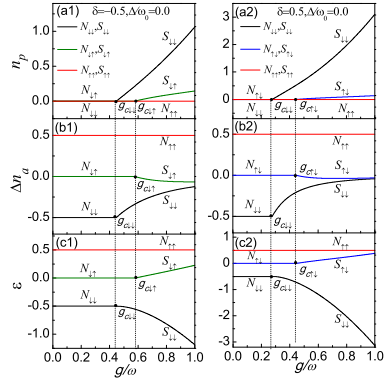}
\caption{The average photon number $n_{p}$ (a), atomic population $\Delta
n_{a}$ (b), and average energy $\protect\varepsilon $ (c) curves for the
imbalance parameter $\protect\delta =-0.5$ (1), $\protect\delta =0.5$ (2) in
the resonance condition $\Delta =0.0$. The stimulated radiation shifts from
one component to the other by adjusting the relative coupling constants.}
\end{figure}

We display the variation curves of average photon-number $n_{p}$\ as
shown in Figs. 6(a1) and 6(a2), atom population imbalance $\Delta
n_{a}$\ as shown in Figs. 6(b1) and 6(b2), and energy $\varepsilon
$\ as depicted in Figs. 6(c1) and 6(c2) with respect the coupling
constant $g$ for the
imbalance parameter $\delta =\pm 0.5$ at the resonance condition $\Delta =0$%
. The QPT from normal state $N_{\downarrow \downarrow }$\ to the
superradiant state $S_{\downarrow \downarrow }$\ takes place at the critical
point $g_{c\downarrow \downarrow }=\sqrt{5}/5=0.447214$. In the case $\delta
=-5$ as depicted in Figs. 6(a1)-6(c1), namely the second component has lower
coupling value, an additional transition appears between the collective
excited-states \textbf{\ }$N_{\downarrow \uparrow }$ and $S_{\downarrow
\uparrow }$ at the critical point $g_{c\downarrow \uparrow }=\sqrt{3}%
/3=0.577350$. This transition is from the normal state of atomic population
inversion to the superradiant state for the second component realized from
atom population imbalance and the energy in Figs. 5(b1) and 5(c1). By
adjusting the imbalance parameter to $\delta =0.5$ as shown Figs.
6(a2)-6(c2), the transition becomes from $N_{\uparrow \downarrow }$\ to $%
S_{\uparrow \downarrow }$\ for the first component. The collective
stimulated-radiation shifts to the first component, which has lower
atom-field coupling than the second component in this case. The transition
critical point is found as $g_{c\uparrow \downarrow }=\sqrt{5}/5=0.447214$.

\section{Conclusion and discussion}

In summary, multiple MQSs are derived analytically for two-component BECs in
a single-mode cavity by means of the spin coherent-state variational method.
The rich phase diagrams are presented with the variation of atom-field
coupling imbalance between two components and the atom-field frequency
detuning. Indeed the ground states display a typical Dicke-model QPT from
the NP to SP for both components in the normal spin-states. When the
atom-field coupling imbalance between two components increases the normal
spin-state with relatively lower coupling-value flips to the inverted
spin-state, the radiation from this state is the stimulated radiation from
atomic population-inversion levels. The stimulated radiation can be also
generated from manipulation of atom-field frequency detuning. In the
specific cases when one of the coupling constants vanishes or two couplings
are equal the ground-states and related QPT reduce to that of an ordinary
Dicke model. The controllable stimulated radiation may have technical
applications in the laser physics. The spin coherent-state variational
method is a powerful tool in the study of macroscopic quantum properties for
the atom-ensemble and cavity-field system, since it takes into account both
the normal and inverted pseudospins, which result in multiple MQSs in
agreement with the semiclassical dynamics of nonequilibrium QPT in the Dicke
model \cite{BMS12}. In addition a one-parameter variational energy-function
is able to be derived in this formalism, so that one can evaluate the
second-order derivative to determine rigorously the local minima of energy
functions in consistence with the numerical simulation \cite%
{WLL16,ABha14,ZLL14,KBS10,NIJL13}.

\section*{Funding}

This work was supported by the National Natural Science Foundation of China
(Grant Nos. 11275118, 11404198, 91430109, 61505100), and the Scientific and
Technological Innovation Programs of Higher Education Institutions in Shanxi
Province (STIP) (Grant No. 2014102), and the Launch of the Scientific
Research of Shanxi University (Grant No. 011151801004), and the National
Fundamental Fund of Personnel Training (Grant No. J1103210). The natural
science foundation of Shanxi Province(Grant No. 2015011008).\


\begin{thebibliography}{99}
\bibitem{Dic54} R. H. Dicke, ``Coherence in Spontaneous Radiation
Processes,'' Phys. Rev. \textbf{93}, 99 (1954).

\bibitem{BGB10} K. Baumann, C. Guerlin, F. Brennecke and T.
Esslinger,\textquotedblleft Dicke quantum phase transition with a superfluid
gas in an optical cavity,\textquotedblright\ Nature (London) \textbf{464},
1301 (2010).

\bibitem{BMB11} K. Baumann, R. Mottl, F. Brennecke and T. Esslinger,
\textquotedblleft Exploring symmetry breaking at the Dicke quantum phase
transition,\textquotedblright\ Phys. Rev. Lett. \textbf{107}, 140402 (2011).

\bibitem{RDB13} H. Ritsch, P. Domokos, F. Brennecke, and T. Esslinger,
\textquotedblleft Cold atoms in cavity-generated dynamical optical
potentials,\textquotedblright\ Rev. Mod. Phys. \textbf{85}, 553 (2013).

\bibitem{WaH73} Y. K. Wang and F. T. Hioe, \textquotedblleft Phase
transition in the Dicke model of superradiance,\textquotedblright\ Phys.
Rev. A, \textbf{7}, 831 (1973).

\bibitem{Hio73} F. T. Hioe, \textquotedblleft Phase transitions in some
generalized Dicke models of superradiance,\textquotedblright\ Phys. Rev. A
\textbf{8}, 1440 (1973).

\bibitem{EmB03} C. Emary and T. Brandes, \textquotedblleft Chaos and the
quantum phase transition in the Dicke model,\textquotedblright\ Phys. Rev. E
\textbf{67}, 066203 (2003).

\bibitem{YCT07} Y. Colombe, T. Steinmetz, G. Dubois, F. Linke, D. Hunger,
and J. Reichel, \textquotedblleft Strong atom--field coupling for
Bose--Einstein condensates in an optical cavity on a
chip,\textquotedblright\ Nature (London) \textbf{450}, 272 (2007).

\bibitem{BDR07} F. Brennecke, T. Donner, S. Ritter, T. Bourdel, M. K\"{o}hl,
and T. Esslinger, \textquotedblleft Cavity QED with a Bose--Einstein
condensate,\textquotedblright\ Nature (London) \textbf{450}, 268 (2007).

\bibitem{RicardoPuebla13} R. Puebla, A. Rela\~{n}o, and J. Retamosa,
\textquotedblleft Excited-state phase transition leading to
symmetry-breaking steady states in the Dicke model,\textquotedblright\ Phys.
Rev. A \textbf{87}, 023819 (2013).

\bibitem{LDM08} J. Larson, B. Damski, G. Morigi, and M. Lewenstein,
\textquotedblleft Mott-insulator states of ultracold atoms in optical
resonators,\textquotedblright\ Phys. Rev. Lett. \textbf{100}, 050401 (2008).

\bibitem{SMorrison08} S. Morrison, and A. S. Parkins, \textquotedblleft
Dynamical quantum phase transitions in the dissipative Lipkin-Meshkov-Glick
model with proposed realization in optical cavity QED,\textquotedblright\
Phys. Rev. Lett. \textbf{100}, 040403 (2008).

\bibitem{JMZhangW08} J. M. Zhang, W. M. Liu, and D. L. Zhou,
\textquotedblleft Mean-field dynamics of a Bose Josephson junction in an
optical cavity,\textquotedblright\ Phys. Rev. A \textbf{78}, 043618 (2008).

\bibitem{CWL08} G. Chen, X. G. Wang, J. -Q. Liang, and Z. D. Wang,
\textquotedblleft Exotic quantum phase transitions in a Bose-Einstein
condensate coupled to an optical cavity,\textquotedblright\ Phys. Rev. A
\textbf{78}, 023634 (2008).

\bibitem{LaM10} J. Larson, and J. -P. Martikainen, \textquotedblleft
Ultracold atoms in a cavity-mediated double-well system,\textquotedblright\
Phys. Rev. A \textbf{82}, 033606 (2010).

\bibitem{ZPL10} L. Zhou, H. Pu, H. Y. Ling, K. Zhang, and W. P. Zhang,
\textquotedblleft Spin dynamics and domain formation of a spinor
Bose-Einstein condensate in an optical cavity,\textquotedblright\ Phys. Rev.
A \textbf{81}, 063641 (2010).

\bibitem{BHS09} M. J. Bhaseen, M. Hohenadler, A. O. Silver, and B. D.
Simons, \textquotedblleft Polaritons and Pairing Phenomena in Bose-Hubbard
Mixtures,\textquotedblright\ Phys. Rev. Lett. \textbf{102}, 135301 (2009).

\bibitem{SHB10} A. O. Silver, M. Hohenadler, M. J. Bhaseen, and B. D.
Simons, \textquotedblleft Bose-Hubbard models coupled to cavity light
fields,\textquotedblright\ Phys. Rev. A \textbf{81}, 023617 (2010).

\bibitem{SND09} G. Szirmai, D. Nagy, and P. Domokos, \textquotedblleft
Excess noise depletion of a Bose-Einstein Condensate in an optical
cavity,\textquotedblright\ Phys. Rev. Lett. \textbf{102}, 080401 (2009).

\bibitem{SND10} G. Szirmai, D. Nagy, and P. Domokos, \textquotedblleft
Quantum noise of a Bose-Einstein condensate in an optical cavity,
correlations, and entanglement,\textquotedblright\ Phys. Rev. A \textbf{81},
043639 (2010).

\bibitem{LLM11} N. Liu, J. L. Lian, J. Ma, L. T. Xiao, G. Chen, J-Q. Liang,
and S. T. Jia, \textquotedblleft Light-shift-induced quantum phase
transitions of a Bose-Einstein condensate in an optical
cavity,\textquotedblright\ Phys. Rev. A \textbf{83}, 033601 (2011).

\bibitem{Tho77} B. V. Thompson, \textquotedblleft A canonical transformation
theory of the generalized Dicke model,\textquotedblright\ J. Phys. A \textbf{%
10}, 89 (1977).

\bibitem{TSo07} D. Tolkunov and D. Solenov, \textquotedblleft Quantum phase
transition in the multimode Dicke model,\textquotedblright\ Phys. Rev. B
\textbf{75}, 024402 (2007).

\bibitem{MDM08} M. Mariantoni, F. Deppe, A. Marx, R. Gross, F. K.
Wilhelm,and E. Solano, \textquotedblleft Two-resonator circuit quantum
electrodynamics: A superconducting quantum switch,\textquotedblright\ Phys.
Rev. B \textbf{78}, 104508 (2008).

\bibitem{NCO09} D. G. Norris, E. J. Cahoon, and L. A. Orozco,
\textquotedblleft Atom detection in a two-mode optical cavity with
intermediate coupling: Autocorrelation studies,\textquotedblright\ Phys.
Rev. A \textbf{80}, 043830 (2009).

\bibitem{TOK09} M. L. Terraciano, R. Olson Knell, D. G. Norris, J. Jing, A.
Fern\'{a}ndez, and L. A. Orozco, \textquotedblleft Photon burst detection of
single atoms in an optical cavity,\textquotedblright\ Nat. Phys. \textbf{5},
480 (2009).

\bibitem{NCi11} P. Nataf and C. Ciuti, \textquotedblleft Protected quantum
computation with multiple resonators in ultrastrong coupling circuit
QED,\textquotedblright\ Phys. Rev. Lett. \textbf{107}, 190402
(2011).

\bibitem{YNo11} J. Q. You and F. Nori, \textquotedblleft Atomic physics and
quantum optics using superconducting circuits,\textquotedblright\ Nature
(London) \textbf{474}, 589 (2011).

\bibitem{WMB11} H. Wang, M. Mariantoni, R. C. Bialczak, M. Lenander, E.
Lucero, M. Neeley, A. D. O'Connell, D. Sank, M. Weides, J. Wenner, T.
Yamamoto, Y. Yin, J. Zhao, J. M. Martinis, and A. N. Cleland,
\textquotedblleft Deterministic entanglement of photons in two
superconducting microwave resonators,\textquotedblright\ Phys. Rev. Lett.
\textbf{106}, 060401 (2011).

\bibitem{ENZ11} Y. Eto, A. Noguchi, P. Zhang, M. Ueda, and M. Kozuma,
\textquotedblleft Projective measurement of a single nuclear spin qubit by
using two-mode cavity QED,\textquotedblright\ Phys. Rev. Lett. \textbf{106},
160501 (2011).

\bibitem{MWB11} M. Mariantoni, H. Wang, R. C. Bialczak, M. Lenander, E.
Lucero, M. Neeley, A. D. O'Connell, D. Sank, M. Weides, J. Wenner, T.
Yamamoto, Y. Yin, J. Zhao, J. M. Martinis, and A. N. Cleland,
\textquotedblleft Photon shell game in three-resonator circuit quantum
electrodynamics,\textquotedblright\ Nat. Phys. \textbf{7}, 287 (2011).

\bibitem{EWi13} D. J. Egger and F. K. Wilhelm, \textquotedblleft Multimode
circuit quantum electrodynamics with hybrid metamaterial transmission
lines,\textquotedblright\ Phys. Rev. Lett. \textbf{111}, 163601 (2013).

\bibitem{YSZ13} C.-P. Yang, Q.-P. Su, S.-B. Zheng, and S. Y. Han,
\textquotedblleft Generating entanglement between microwave photons and
qubits in multiple cavities coupled by a superconducting
qutrit,\textquotedblright\ Phys. Rev. A \textbf{87}, 022320 (2013).

\bibitem{LLe09} J. Larson and S. Levin, \textquotedblleft Effective abelian
and non-abelian gauge potentials in cavity QED,\textquotedblright\ Phys.
Rev. Lett. \textbf{103}, 013602 (2009).

\bibitem{Lar10} J. Larson, \textquotedblleft Analog of the
spin-orbit-induced anomalous Hall effect with quantized
radiation,\textquotedblright\ Phys. Rev. A \textbf{81}, 051803 (2010).

\bibitem{GLG10} S. Gopalakrishnan, B. L. Lev, and P. M.
Goldbart,\textquotedblleft Emergent crystallinity and frustration
with Bose-Einstein condensates in multimode
cavities,\textquotedblright\ Nat. Phys. \textbf{5}, 845 (2009);
\textquotedblleft Atom-light crystallization of Bose-Einstein
condensates in multimode cavities: Nonequilibrium classical and
quantum phase transitions, emergent lattices, supersolidity, and
frustration,\textquotedblright\ Phys. Rev. A \textbf{82}, 043612
(2010).

\bibitem{GLG11} S. Gopalakrishnan, B. L. Lev, and P. M. Goldbart,
\textquotedblleft Frustration and glassiness in spin models with
cavity-mediated interactions,\textquotedblright\ Phys. Rev. Lett. \textbf{107%
}, 277201 (2011).

\bibitem{SSa11} P. Strack and S. Sachdev, \textquotedblleft Dicke quantum
spin glass of atoms and photons,\textquotedblright\ Phys. Rev. Lett. \textbf{%
107}, 277202 (2011).

\bibitem{BSS13} M. Buchhold, P. Strack, S. Sachdev, and S. Diehl,
\textquotedblleft Dicke-model quantum spin and photon glass in optical
cavities: Nonequilibrium theory and experimental
signatures,\textquotedblright\ Phys. Rev. A \textbf{87}, 063622 (2013).

\bibitem{FYZ14} J. T. Fan, Z. W. Yang, Y. W. Zhang, J. Ma, G. Chen and S. T.
Jia,\textquotedblleft Hidden continuous symmetry and Nambu-Goldstone mode in
a two-mode Dicke model,\textquotedblright\ Phys. Rev. A, \textbf{89}, 023812
(2014).

\bibitem{WHR13} A. Wickenbroc, M. Hemmerling, G. R. M. Robb, C. Emary, and
F. Renzoni, \textquotedblleft Collective strong coupling in multimode cavity
QED,\textquotedblright\ Phys. Rev. A \textbf{87}, 043817 (2013).

\bibitem{KLR} D. O. Krimer, M. Liertzer, S. Rotter, and H. E. T\"{u}%
reci,\textquotedblleft Route from spontaneous decay to complex multimode
dynamics in cavity QED,\textquotedblright\ arXiv:1306.4787.

\bibitem{KV08} T. J. Kippenberg, and K. J. Vahala, \textquotedblleft Cavity
optomechanics: Back-action at the mesoscale,\textquotedblright\ Science
\textbf{321}, 1172 (2008)

\bibitem{MG09} F. Marquardt, and S. M. Girvin,\textquotedblleft
Optomechanics,\textquotedblright\ Physics \textbf{2}, 40 (2009).

\bibitem{FK09} I. Favero, and K. Karrai,\textquotedblleft optomechanics of
deformable optical cavities,\textquotedblright\ Nature Photonics \textbf{3},
201 (2009).

\bibitem{AGHK10} M. Aspelmeyer, S. Gr\"{o}blacher, K. Hammerer, and N.
Kiesel, \textquotedblleft Quantum optomechanics---throwing a
glance,\textquotedblright\ J. Opt. Soc. Am. B \textbf{27}, A 189
(2010).

\bibitem{HWAH11} C. A. Regal and K. W. Lehnert, \textquotedblleft From
cavity electromechanics to cavity optomechanics,\textquotedblright\ J.
Phys.: Conf. Ser.\textbf{\ 264}, 012025 (2011).

\bibitem{WLL16} Z. M. Wang, J. L. Lian, J.-Q. Liang, Y. M. Yu, and W. M.
Liu,\textquotedblleft Collapse of the superradiant phase and multiple
quantum phase transitions for Bose-Einstein condensates in an optomechanical
cavity,\textquotedblright\ Phys. Rev. A \textbf{93}, 033630 (2016).

\bibitem{LiangJQ} J.-Q. Liang, J.-L. Liu, W.-D. Li, and Z.-J.
Li,\textquotedblleft Atom-pair tunneling and quantum phase
transition in the strong-interaction regime,\textquotedblright\
Phys. Rev. A \textbf{79}, 033617 (2009).

\bibitem{FuLB} H. Cao and L. B. Fu,\textquotedblleft Quantum phase
transition and dynamics induced by atom-pair tunnelling of
Bose-Einstein condensates in a double-well
potential,\textquotedblright\ Eur. Phys. J. D \textbf{66}, 97
(2012).

\bibitem{ZhangYC} Y. C. Zhang, X. F. Zhou, G. C. Guo, X. X. Zhou, H. Pu and
Z. W. Zhou, \textquotedblleft Two-component polariton condensate in
an optical microcavity,\textquotedblright\ Phys. Rev. A \textbf{89},
053624 (2014).

\bibitem{TimmermansE} E. Timmermans, \textquotedblleft Phase separation of
Bose-Einstein condensates,\textquotedblright\ Phys. Rev. Lett.
\textbf{81}, 5718 (1998).

\bibitem{PuH} H. Pu and N.P. Bigelow, \textquotedblleft Properties of
two-species Bose condensates,\textquotedblright Phys. Rev. Lett.
\textbf{80}, 1130 (1998).

\bibitem{DongY} Y. Dong, J. W. Ye, and H. Pu, \textquotedblleft
Multistability in an optomechanical system with a two-component
Bose-Einstein condensate,\textquotedblright\ Phys. Rev. A
\textbf{83}, 031608(R) (2011).

\bibitem{SasakiK} K Sasaki, N. Suzuki, and H. Saito,\textquotedblleft
Capillary instability in a two-component Bose-Einstein
condensate,\textquotedblright\ Phys. Rev. A \textbf{83}, 053606
(2011).

\bibitem{BarankovRA} R. A. Barankov, \textquotedblleft Boundary of two mixed
Bose-Einstein condensates,\textquotedblright\ Phys. Rev. A
\textbf{66}, 013612 (2002).

\bibitem{SchaeybroeckBV} B. V. Schaeybroeck, \textquotedblleft Interface
tension of Bose-Einstein condensates,\textquotedblright\ Phys. Rev.
A \textbf{78}, 023624 (2008); \textquotedblleft Addendum to
\textquotedblleft Interface tension of Bose-Einstein
condensates,\textquotedblright\ Phys. Rev. A \textbf{80}, 065601
(2009).

\bibitem{Sasakik} K. Sasaki, N. Suzuki, D. Akamatsu, and H.
Saito,\textquotedblleft Rayleigh-Taylor instability and
mushroom-pattern formation in a two-component Bose-Einstein
condensate,\textquotedblright\ Phys. Rev. A \textbf{80}, 063611
(2009).

\bibitem{SorensenA} A. S\o ensen, L.-M. Duan, J. I. Cirac, and P. Zoller,
\textquotedblleft Many-particle entanglement with Bose-Einstein
condensates,\textquotedblright\ Nature(London) \textbf{409}, 63
(2001).

\bibitem{GordonD} D. Gordon and C. M. Savage, \textquotedblleft Creating
macroscopic quantum superpositions with Bose-Einstein
condensates,\textquotedblright Phys. Rev. A \textbf{59}, 4623
(1999).

\bibitem{MicheliA} A. Micheli, D. Jaksch, J. I. Cirac, and P.
Zoller,\textquotedblleft Many-particle entanglement in two-component
Bose-Einstein condensates,\textquotedblright\ Phys. Rev. A
\textbf{67}, 013607 (2003).

\bibitem{AndrewsMR} M. R. Andrews, C. G. Townsend, H.-J. Miesner, D. S.
Durfee, D. M. Kurn, W. Ketterle,\textquotedblleft Observation of
interference between two Bose condensates,\textquotedblright\
Science \textbf{275}, 637 (1997).

\bibitem{EtoM} M. Eto, K. Kasamatsu, M. Nitta, H. Takeuchi, and M. Tsubota,
\textquotedblleft Interaction of half-quantized vortices in
two-component Bose-Einstein condensates,\textquotedblright\ Phys.
Rev. A \textbf{83}, 063603 (2011).

\bibitem{EmaryC} C. Emary, and T. Brandes, \textquotedblleft Quantum chaos
triggered by precursors of a quantum phase transition: The Dicke
model,\textquotedblright\ Phys. Rev. Lett. \textbf{90}, 044101
(2003).

\bibitem{ChenG} G. Chen, J. Q. Li, and J.-Q. Liang, \textquotedblleft
Critical property of the geometric phase in the Dicke
model,\textquotedblright\ Phys. Rev. A \textbf{74}, 054101 (2006).

\bibitem{Lian1} J. L. Lian, Y. W. Zhang, and J.-Q Liang, \textquotedblleft
Macroscopic quantum states and quantum phase transition in the Dicke
model ,\textquotedblright Chin. Phys. Lett. \textbf{29}, 060302
(2012).

\bibitem{Lian2} J. L. Lian, N. Liu, J.-Q Liang, G. Chen, and S. T.
Jia,\textquotedblleft Ground-state properties of a Bose-Einstein
condensate in an optomechanical cavity,\textquotedblright\ Phys.
Rev. A \textbf{88}, 043820 (2013).

\bibitem{NIJL13} N. Liu, J. D. Li, and J. -Q. Liang,\textbf{%
\textquotedblleft }Nonequilibrium quantum phase transition of Bose-Einstein
condensates in an optical cavity\textbf{,\textquotedblright\ }Phys. Rev. A
\textbf{87}, 053623 (2013).

\bibitem{GilmoreR} R. Gilmore, L.M. Narducci,\textquotedblleft Relation
between the equilibrium and nonequilibrium critical properties of
the Dicke model,\textquotedblright\ Phys. Rev. A \textbf{17}, 1747
(1978).

\bibitem{DimerF} F. Dimer, B. Estienne, A. S. Parkins, and H. J. Carmichael,
\textquotedblleft Proposed realization of the Dicke-model quantum
phase transition in an optical cavity QED system,\textquotedblright\
Phys. Rev. A \textbf{75}, 013804 (2007).

\bibitem{HorakP} P. Horak, and H. Ritsch, \textquotedblleft Dissipative
dynamics of Bose condensates in optical cavities,\textquotedblright\
Phys. Rev. A \textbf{63}, 023603 (2001).

\bibitem{CastannosO} O. Casta\~{n}os, E. Nahmad-Achar, R. L\'{o}pez-Pe\~{n}%
na, and J. G. Hirsch, \textquotedblleft No singularities in
observables at the phase transition in the Dicke
model,\textquotedblright\ Phys. Rev. A
\textbf{83}, 051601(R)
(2011).

\bibitem{ZLL14} X. Q. Zhao, N. Liu and J.-Q. Liang,\textquotedblleft
Nonlinear atom-photon-interaction-induced population inversion and inverted
quantum phase transition of Bose-Einstein condensate in an optical
cavity,\textquotedblright\ Phys. Rev. A \textbf{90}, 023622 (2014).

\bibitem{CLS04} Z.-D. Chen, J.-Q. Liang, S.-Q. Shen, and W. -F. Xie,
\textquotedblleft Dynamics and Berry phase of two-species Bose-Einstein
condensates,\textquotedblright\ Phys. Rev. A \textbf{69}, 023611 (2004).

\bibitem{KBS10} J. Keeling, M. J. Bhaseen, and B. D. Simons,
\textquotedblleft Collective dynamics of Bose-Einstein condensates in
optical cavities,\textquotedblright\ Phys. Rev. Lett. \textbf{105}, 043001
(2010).

\bibitem{BMS12} M. J. Bhaseen, J. Mayoh, B. D. Simons and J. Keeling,
\textquotedblleft Dynamics of nonequilibrium Dicke
models,\textquotedblright\ Phys. Rev. A, \textbf{85}, 013817 (2012).

\bibitem{ABha14} A. B. Bhattacherjee, \textquotedblleft Non-equilibrium
dynamical phases of two-Atom Dicke model,\textquotedblright\ Phys. Lett. A
\textbf{378}, 3244 (2014).

\bibitem{LHC90} E. Layton, Y. H. Huang, and S. I. Chu, \textquotedblleft
Cyclic quantum evolution and Aharonov-Anantlan geometric phases in SU(2)
spin-coherent states,\textquotedblright\ Phys. Rev. A \textbf{41}, 42 (1990).

\bibitem{RFF99} R. F. Fox, \textquotedblleft Generalized coherent
states,\textquotedblright\ Phys. Rev. A \textbf{59}, 3241 (1999).

\bibitem{YZLLM96} Y.-Z. Lai, J.-Q. Liang, H. J. W. M\"{u}ller-Kirsten, and
J. G. Zhou, \textquotedblleft Time-dependent quantum systems and the
invariant Hermitian operator,\textquotedblright\ Phys. Rev. A \textbf{53},
3691 (1996).
\end{thebibliography}
\end{document}